\documentclass[aip,amsmath,amssymb,preprint]{revtex4-1}
\usepackage{graphicx}
\usepackage{dcolumn}
\usepackage{bm}

\usepackage[utf8]{inputenc}
\usepackage[T1]{fontenc}
\usepackage{mathptmx}

\usepackage{amsbsy,amssymb,amsfonts}
\usepackage{color}
\newcommand{\beq}{\begin{equation}}
\newcommand{\eeq}{\end{equation}}
\newcommand{\beqa}{\begin{eqnarray}}
\newcommand{\eeqa}{\end{eqnarray}}

\newcommand{\bef}{\begin{figure}}
\newcommand{\ef}{\end{figure}}
\newcommand{\bc}{\begin{center}}
\newcommand{\ec}{\end{center}}
\newcommand{\bt}{\begin{table}}
\newcommand{\et}{\end{table}}
\newcommand{\btb}{\begin{tabular}}
\newcommand{\etb}{\end{tabular}}

\newcommand{\maps}{{\emph{a priori }}}
\newcommand{\mapos}{{\emph{a posteriori }}}

\begin{document}


\title {Transformation to a geminal basis and stationary conditions for the exact wave function therein}

\author{Lasse K. S{\o}rensen}
\affiliation{University Library, University of Southern Denmark, DK-5230 Odense M, Denmark}
\email{lasse.kragh.soerensen@gmail.com}

\date{\today}
\begin{abstract}
We show the transformation from a one-particle basis to a geminal basis, transformations between different geminal bases and demonstrate the Lie algebra of a geminal basis. From the basis transformations we express both the wave function and Hamiltonian in the geminal basis.  
The necessary and sufficient conditions of the exact wave function expanded in a geminal basis is shown to be a Brillouin theorem of geminals. The variational optimization of the geminals in the Antisymmetrized Geminal Power (AGP), Antisymmetrized Product of Geminals (APG) and the Full Geminal Product (FGP) wave function ans{\"a}tze are discussed. We show that using a geminal replacement operator to describe geminal rotations introduce both primary and secondary rotations. The secondary rotations rotate two geminals in the reference at the same time due to the composite boson nature of geminals. 
Due to the completeness of the FGP, where all possible geminal combinations are present, the FGP is exact. The number of parameters in the FGP scale exponentially with the number of particles, like the Full Configuration Interaction (FCI). Truncation in the FGP expansion could lead to compact representations of the wave function in the future since the reference function in the FGP is the APG wave function.

\end{abstract}

\maketitle

\section{Introduction}
The formulation of compact wave functions, be it exact or approximate formulations, are a central part of both quantum physics and chemistry. The guiding principles behind these formulations are often vague though attempts at stricter measures of compactness from entropy measures are currently being developed.\cite{renyi_entropy,FLORESGALLEGOS201861,Shannon_entropy_ci,info_tech,shannon_entropy} While correlation between entropy and compactness is seen for the methods investigated with the entropy measures these methods are, unlike full configuration interaction (FCI), not invariant to orbital rotations and the result may therefore be dependent on the choice of orbitals. The usefulness of the entropy as a black box measure of compactness of a wave function is therefore not straight forward.
The more traditional way of analysing the quality and compactness of a wave function is the statistical analysis of direct numerical comparisons.\cite{olsen_num_comp,Puzzarini_2016} While the statistical analysis is straight forward it is, however, difficult to extend to large systems since there is no exact result to compare against and the comparisons are therefore limited in the number of electrons and basis functions.

Creating compact representations of the wave function, expanded in a one-particle basis set, which capture both static and dynamic correlation has proven to be challenging. The classic methods such as Multi-Configurational Self Consistent Field (MCSSCF) and the specific type of MCSCF namely complete active space self-consistent field (CASSCF) and the more modern methods such as density matrix renormalization group (DMRG) and diffusion quantum Monte Carlo (DQMC) all have proven very good in capturing the static correlation with system sizes up to around 100 orbitals for the modern methods.\cite{mcscf_review_shepard,roos_casscf1,White_DMRG,GRIMM1971134} For the dynamic correlation of MCSCF, CASSCF, DMRG and DQMC these methods often rely on second order perturbation theory.\cite{caspt2,DMRG_review_reiher,qmc_pt_blunt} The straight forward application of perturbation theory, which require higher-order reduced density matrices, on these large active spaces unfortunately makes the perturbation theory the bottleneck. 
For methods like multi-reference configuration interaction (MRCI) and the many flavours of multi-reference coupled cluster (MRCC) where both static and dynamic correlation is treated simultaneous very accurate calculations on larger systems is hampered by the rapid scaling increase with every improvement in the correlation level. 

Wave function ans{\"a}tze based on geminals, which are two-electron functions, present a promising alternative to conventional state-of-the-art electronic structure methods to model both static and dynamic correlation. Tecmer and Boguslawski recently reviewed the current progress in creating gemninal ans{\"a}tze to deal with strong correlation, missing dynamical correlation, correlation extensions, excites states and open shells.\cite{geminal_review_tecmer} The review clearly demonstrates that geminal wave function methods are a growing niche. We will here not focus on any of these many interesting extensions but more on the Lie algebra of geminals, the transformation from the one-particle basis to the geminal basis and working consistently in a geminal basis.

Recently it was shown that the necessary and sufficient conditions for the exact wave function always can be written as a generalized Brillouin theorem irrespective of the order of interaction in the Hamiltonian.\cite{Lasse_Brillouin} For a Hamiltonian containing only one-electron terms the exact wave function can be found from the Brillouin theorem by minimizing the energy with respect to rotations between the occupied and virtual orbitals in a one-particle basis set.\cite{thouless_theorem}
We will here formulate the necessary and sufficient conditions, or more aptly the stationary conditions, for the exact wave function in a geminal basis.
From the stationary conditions we examine the optimization effect of geminal rotations and geminal replacement operator for different geminal wave function ans\"atze.

The variational optimization using geminal rotations and geminal replacement operators of the two simplest geminal wave function ans{\"a}tze, which are the Antisymmetrized Geminal Power (AGP)\cite{superconductivity,coleman1964_super,coleman1964_super2,coleman1965,coleman1997} and the Antisymmetrized Product of Geminals (APG),\cite{kurtz_gagp,silvern(n-1)/2,silver2,nicely1,Dyson1956,JANSSEN1971145,10.1143/PTP.57.1554,PhysRevC.52.1394,DOBACZEWSKI1981213,RevModPhys.63.375,RevModPhys.84.711}  will be examined. The Full Geminal Product (FGP), which is a linear combinations of all possible geminal products, is the equivalent of the Full Configuration Interaction (FCI) in a geminal basis. We will here show that the FGP is exact by virtue of completeness of the expansion, exactly like the FCI in a one-particle basis set.

We will throughout use $p,q,\ldots$ and $\mu , \nu,\ldots $ as general indices and $a, b, \ldots$ and $\alpha, \beta, \ldots$ as occupied indices when referring to orbital and geminal indices, respectively, unless stated otherwise. We will here use real orbitals and a real coefficient matrix but this is straight forward to extend to include complex algebra.


\section{Geminal basis and algebra}
\label{sec:gemalg}

In order to obtain a proper pair rotation a geminal basis is needed since the general pair rotation in a one-particle basis does not fulfill a Lie algebra.\cite{muk_kut_gcc,kut_muk_min_par} The geminal basis have been explored rather extensively in quantum chemistry though always in a very approximate form.\cite{kurtz_gagp,coleman1965,coleman1997,silvern(n-1)/2,silver2,nicely1,apsg1,apsg2} In nuclear physics a restricted geminal basis or pair fermion basis have been applied extensively in Dyson or other boson-fermion type mappings.\cite{Dyson1956,BELIAEV1962582,JANSSEN1971145,10.1143/PTP.57.1554,PhysRevC.52.1394,DOBACZEWSKI1981213,RevModPhys.63.375,RevModPhys.84.711} 

In this section we will review the general geminal algebra where the aim is to show the Lie algebra and how the structure constants from any geminal resulting from nested commutators easily can be found by projecting onto the geminal basis. Secondly we will write the Hamiltonian in a geminal basis and show that the necessary and sufficient conditions for the exact wave function in a geminal basis is a generalization of the known Brillouin theorem. 

The notation from Surj$\rm \acute{a}$n, where $\Psi^+_{\mu}$ and $\Psi^-_{\nu}$ are the general geminal creation and annihilation of geminal $\mu$ and $\nu$, respectively, will be used.\cite{surjan_review,surjan_book} Additional simple geminal relations are listed in Appendix \ref{gemalg}. A recap of the necessary and sufficient conditions for the exact wave function is given in Appendix \ref{sec:necessary}. The derivation of basis set transformations in a geminal basis along with the transition from a one-electron basis to a geminal basis are shown in Appendix \ref{appendix_ham} along with the transformation to a natural geminal.

\subsection{The geminal basis}
\label{sec:defgem}

In a spin-orbital basis the most general way of writing a geminal $\Psi^{+}_{\mu}$ is
\beq
\label{geminal}
\Psi_{\mu}^{+} =  \frac{1}{2} \sum_{pq, \sigma \upsilon}^{m,s} C_{p\sigma q\upsilon} a^{\dagger}_{p \sigma} a^{\dagger}_{q \upsilon}
\eeq
where $C_{p\sigma q\upsilon}$ is an element in a skew-symmetric coefficient matrix and the sum is over all orbitals $m$ and their respective spin-functions $s$.
The annihilation geminal is taken as the Hermitian adjoint of $\Psi^{+}_{\mu}$
\beq
\label{geminala}
\Psi_{\mu}^{-} =  \frac{1}{2} \sum_{pq, \sigma \upsilon}^{m,s} C_{p\sigma q\upsilon} a_{q \upsilon} a_{p \sigma}.
\eeq

The natural form of a geminal is often used due to the simplicity in the diagonal of the natural geminal. Since we here are interested in expressing any wave function and Hamiltonian in a complete gemnial basis the natural geminal is here not of interest since in it is not possible to transform all geminals in the basis to the natural form at the same time as discussed in Appendix \ref{ssec:natural}.  

\subsection{Normalization}

Exactly like orbitals, which are able to form an orthonormal set, we will likewise demand that the geminals form an orthonormal set
\beq
\label{gort}
\langle vac | \Psi_{\nu}^- \Psi_{\mu}^+ | vac \rangle = \delta_{\nu \mu}.
\eeq
The orthonormality condition in Eq. \ref{gort} is usually, in quantum chemistry, referred to as the weak orthogonality condition and is just the Frobenius inner product of the coefficient matrices
\beq
\frac{1}{2} tr(\mathbf{C}^{T} \mathbf{C}) = 1.
\eeq
The Frobenius inner product is a component-wise inner product of two matrices which treat the matrices as vectors with an inner product.

In quantum chemistry it has been customary to use the strong orthogonality instead of the weak orthogonality. While the strong orthogonality condition simplifies the geminal algebra,\cite{apsg1,apsg2} due to Arai's theorem where  where an orbital will only belong to a single geminal,\cite{arai_theorem,arai_theorem2,inge1,inge7} we will here focus on the general case where the weak orthogonality is the only constraint on the geminal basis.

\subsection{Lie algebra and commutation relations}
\label{sec:superalg}

We will here present the most important geminal algebra. A few more simple relations are also summarized in Appendix \ref{gemalg} and the rotations and transformations of geminals are shown in Appendix \ref{appendix_ham}. Since the geminals are composite bosons they do not form a nice algebra like real bosons\cite{surjan_book,surjan_review,kvasnicka} and show neither real Bose-Einstein nor Fermi-Dirac statistics\cite{coleman1964_super,coleman1964_super2}.

The commutation of the creation and annihilation geminal operators among them selves follows that of regular bosons
\beq
\label{commutation_op}
[ \Psi_{\nu}^+ , \Psi_{\mu}^+ ] =  
[ \Psi_{\nu}^- , \Psi_{\mu}^- ] = 0.
\eeq
The commutation between the creation and annihilation geminals produce an additional term $R$
\beq
[ \Psi_{\nu}^- , \Psi_{\mu}^+] =
Q_{\nu\mu} =
\delta_{\nu \mu} + R_{\nu\mu}          
\eeq
which is non-zero and need not commute with any other operator. The central problem in constructing a simpler geminal algebra is therefore finding a suitable $R$. Much work has focused on eliminating $R$ since this restores the regular boson algebra. Eliminating $R$ unfortunately only seems possible when using Arai's theorem for the strong orthogonality condition.\cite{arai_theorem,arai_theorem2} 
We will not introduce any attempts at making the algebra simpler but simply show the consequences of the regular geminal algebra.

The commutator between $\Psi_{\nu}^{-}$ and $\Psi_{\mu}^{+}$ illustrates the geminal algebra
\beqa
[ \Psi_{\nu}^{-} , \Psi_{\mu}^{+}] &=&
[ \frac{1}{2} \sum_{pq} C_{pq}^{\nu} a_q a_p , \frac{1}{2} \sum_{rs} C_{rs}^{\mu} a^{\dagger}_r a^{\dagger}_s ]  \nonumber \\
&=& \frac{1}{4} \sum_{pqrs} C_{pq}^{\nu} C_{rs}^{\mu} (
\delta_{qs} \delta_{pr} - \delta_{qr} \delta_{ps} 
+ a^{\dagger}_s a_p \delta_{qr} - a^{\dagger}_r a_p \delta_{qs}
+ a^{\dagger}_r a_q \delta_{ps} - a^{\dagger}_s a_q \delta_{pr}) \nonumber \\
&=& Q_{\nu \mu} =
\delta_{\mu \nu} + R_{\nu \mu}        
\eeqa
where we can use the skew-symmetry of the coefficient matrix to simplify $R$
\beq
\label{r_simple}
R_{\nu \mu} = \sum_{pqs} C_{pq}^{\nu} C_{qs}^{\mu} a^{\dagger}_s a_p.
\eeq

Since $R$ is non-zero it is 
therefore of interest to investigate the 
commutator between $\Psi^{+}$ and $R$
\beqa
\label{superalg}
[[ \Psi_{\nu}^{-} , \Psi_{\mu}^{+}], \Psi_{\gamma}^{+}] &=&
[ Q_{\nu \mu}, \Psi_{\gamma}^{+}] =
[ R_{\nu \mu}, \Psi_{\gamma}^{+}] \nonumber \\
&=& \frac{1}{8} \sum_{pqrstu}   C_{pq}^{\nu}  C_{rs}^{\mu}  
C_{tu}^{\gamma} * \nonumber \\ & &( 
a^{\dagger}_s a^{\dagger}_u (
\delta_{pt}\delta_{qr} - \delta_{pr}\delta_{qt}) +
a^{\dagger}_s a^{\dagger}_t ( \delta_{pr}\delta_{qu} - \delta_{pu}\delta_{qr}) \nonumber \\ & & +
a^{\dagger}_r a^{\dagger}_t ( \delta_{pu}\delta_{qs} - \delta_{ps}\delta_{qu}) +
a^{\dagger}_r a^{\dagger}_u ( \delta_{ps}\delta_{qt} - \delta_{pt}\delta_{qs}) ) \nonumber \\
&=& -\sum_{pqsu}  C_{pq}^{\nu}  C_{sq}^{\mu}  C_{pu}^{\gamma}
a^{\dagger}_s a^{\dagger}_u \nonumber \\
&=& \sum_{\tau} c^{\mu \gamma}_{\nu \tau}  \Psi_{\tau}^{+} = 
\Psi_{\nu \mu \gamma}^{+}.
\eeqa
The commutator gives a new geminal $\Psi_{\nu \mu \gamma}^{+}$ where $c^{\mu \gamma}_{\nu \tau}$ are the structure constants for the expansion in the geminal basis. $\Psi_{\nu \mu \gamma}^{+}$ can be expanded exactly in the geminal when all possible geminals created from $m$ orbitals are included in the geminal basis. The structure constants can easily be found by applying the Frobenius inner product since the Frobenius inner product is a component-wise inner product of two matrices. A similar relation can be derived from the pair annihilator from Eq. \ref{geminala}
\beqa
[[ \Psi_{\nu}^{+} , \Psi_{\mu}^{-} ], \Psi_{\gamma}^{-}] &=& 
[ Q_{\nu \mu}, \Psi_{\gamma}^{-} ] \nonumber \\
\label{superalg2}
&=&
\sum_{\tau} c_{\mu \gamma}^{\nu \tau} \Psi_{\tau}^{-} =
\Psi_{\nu \mu \gamma}^{-} .
\eeqa
As seen from Eqs. \ref{superalg} and \ref{superalg2} the geminals form a Lie algebra.\cite{RevModPhys.63.375,10.1143/PTP.57.1554}
The structure constants will show the following symmetry\cite{PhysRevC.44.1030}
\beq
c^{\mu \gamma}_{\nu \tau} = c^{\gamma \mu}_{\nu \tau} =
c^{\mu \gamma}_{\tau \nu} = (c_{\mu \gamma}^{\nu \tau})^* .
\eeq

The nested commutator for a single geminal, which could be used for internal occupied rotations,
\beq
\label{new_con}
[[ \Psi_{\mu}^{-} , \Psi_{\mu}^{+}], \Psi_{\mu}^{+}] = 
\sum_{\tau} c^{\mu \mu}_{\mu \tau}  \Psi_{\tau}^{+}
\eeq
does not give great simplifications to Eq. \ref{superalg} and shows that even internal rotations of geminals produce additional terms. 

\subsection{Hamiltonian in a geminal basis}

In Appendix \ref{appendix_ham} orbital and geminal rotations along with basis set transformations are presented. From the geminal rotations in Appendix \ref{appendix_ham} it is shown that the transformation from the Hamiltonian in Eq. \ref{hamiltonian} in a one-particle basis lead to a Hamiltonian in a geminal written as
\beq
\label{H_in_gem_basis_a}
\hat H = \sum_{\mu \nu} \omega_{\mu \nu} \Psi_{\mu}^+ \Psi_{\nu}^-.
\eeq
The necessary and sufficient conditions for the exact wave function in a geminal basis therefore reads
\beq
\label{double_g}
0 = \langle \Psi | (\hat H - E) \omega_{\mu \nu} \Psi_{\mu}^+ \Psi_{\nu}^- |  \Psi \rangle 
\eeq
which is seen to be identical to the necessary and sufficient conditions for the exact wave function in a single-particle basis, in a rotated basis, as seen in Eq. \ref{double}. The stationary conditions for the wave function, in a geminal bases, are therefore a Brillouin's theorem of geminals.\cite{Lasse_Brillouin}

\section{Geminal wave function ans{\"a}tze}
\label{geman}

The two simplest geminal ans{\"a}tze, namely the AGP and APG, and a full geminal product (FGP) ansatz will be investigated separately. Both the AGP and APG, when expanded in single-particle functions, gives all configurations in the FCI expansion where the weight of all the configurations can be written as a product of sums. The FGP ansatz is a linear combination of all possible tensor products of geminals.

While the geminals used in the AGP and APG formally are the same the constraints on the geminals may be very different in order to represent a proper N-particle wave function. The FGP on the other hand is invariant to geminal rotations. We will, however, not dwell upon these differences but instead focus on the algebra and geminal rotations.

\subsection{Wave function requirements}

In the AGP and APG it is the geminals in the reference that is being optimized and the requirements are therefore similar to the  requirements seen for a Hamiltonian with only one-body interactions where the orbitals in a configuration-state function (CSF) is being optimized. For the FGP the requirements are on the coefficients in front of every geminal product similar to FCI.

Variations of the reference $\Psi_{0}$ should ideally give
\beq
\label{fog}
\Psi_{0} \rightarrow \Psi_{0} + \delta \Psi_0 = \Psi_0 + \Psi_0 (\Psi_{\alpha}^+ \rightarrow \sum_{\mu} \eta_{\mu \alpha} \Psi_{\mu}^+ ) + \mathcal{O}_2 + \ldots
\eeq
a first order linearly independent variation $\delta \Psi$ to $\Psi_{0}$ where the geminal $\Psi_{\alpha}^+$ in $\Psi_{0}$ is replaced by a sum over of geminals with some expansion coefficient $\eta_{\mu \alpha}$, as indicated by $\Psi_0 (\Psi_{\alpha}^+ \rightarrow \sum_{\mu} \eta_{\mu \alpha} \Psi_{\mu}^+ )$, along with some higher order corrections in $\eta_{\mu \alpha}$ which disappear when the function is optimized. The energy is in this way variational
\beq
\label{variation}
\langle \Psi_{0} | (\hat H - E) | \delta \Psi_0 \rangle = 0
\eeq
and the first derivative with respect to the variational coefficients $\eta_{\mu \alpha}$ should, if the wave function is exact, reproduce the stationary conditions from the necessary and sufficient conditions of the exact wave function.

\subsubsection{The necessary and sufficient conditions of the exact wave function}

The initial conditions for the derivative from Nakatsuji\cite{apnak1}
\beq
\label{strict_nak}
\left. \frac{\partial \Psi_{0}}{\partial C_{qs}^{pr}} \right|_{C =0} = \omega_{pqrs} a_p^{\dagger} a_r^{\dagger} a_s a_q | \Psi_{0} \rangle
\eeq
can be replaced by a weaker conditions as shown by Mukherjee and Kutzelnigg\cite{muk_kut_gcc}
\beq
\label{derivative_loose}
\left. \frac{\partial \Psi_{0}}{\partial C_{qs}^{pr}} \right|_{C =0} = \sum_{p>r,s>q} \omega_{pqrs} a_p^{\dagger} a_r^{\dagger} a_s a_q | \Psi_{0} \rangle.
\eeq


The stricter conditions for a wave function in a geminal basis, as seen from Eq. \ref{double_g}, is
\beq
\label{strict_gem}
\left. \frac{\partial \Psi_{0}}{\partial \eta_{\mu \nu}} \right|_{\eta =0} = \omega_{\mu \nu} \Psi_{\nu}^+ \Psi_{\mu}^- | \Psi_{0} \rangle
\eeq
and following Eq. \ref{derivative_loose} the weaker conditions in a geminal basis is
\beq
\label{loose_gem2}
\left. \frac{\partial \Psi_{0}}{\partial \eta_{\mu \nu}} \right|_{\eta =0} = \sum_{\mu \nu} \omega_{\mu \nu} \Psi_{\nu}^+ \Psi_{\mu}^- | \Psi_{0} \rangle.
\eeq
We will here use the stricter conditions in Eq. \ref{strict_gem} when relating the replacement operators from the derivative to geminal rotations.
The stricter conditions is the geminal equivalent of Brillouin's theorem where the replacement operator of geminals works on the reference.

\subsection{The antisymmetrized geminal power (AGP)}
\label{varagp}

The AGP\cite{superconductivity,coleman1963,coleman1965} have been used successfully for the properties of the solid state and in cooperative phenomena such as for superconductivity and superfluidity in BCS theory\cite{superconductivity,coleman1964_super,coleman1964_super2} but is much less explored for molecules though recently there has been a renewed interest in combining the AGP with CI.\cite{osamu_2015,osamu_2016,scuseria_2019,scuseria_2019_2} 

The AGP is a simple tensor product of of $N/2$ identical geminals
\beq
\label{agp}
\Psi_{AGP} = \prod^{N/2} \Psi_{\alpha}^+ | vac \rangle = 
(\Psi_{\alpha}^+)^{N/2} | vac \rangle
\eeq
where $N$ is the number of electrons, $N$ is here assumed even. 

\subsubsection{Variations of the geminals in the AGP}

Varying the geminal in the AGP
\beqa
\Psi_{AGP} &=& \prod^{N/2} \tilde \Psi_{\alpha}^+ | vac \rangle \nonumber \\
\label{AGP_ansatz}
           &=& \prod^{N/2} ( \Psi_{\alpha}^+ + \sum_{\nu\neq \alpha} \eta_{\nu \alpha} \Psi_{\nu}^+) | vac \rangle
\eeqa
gives the AGP wave function in the transformed geminal basis $\tilde \Psi_{\alpha}^+$. Taking the first derivative of the transformed wave function with respect to the variational coefficients $\eta_{\nu \alpha}$ should reproduce the stationary conditions
\beq
\label{1dev}
\left. \frac{\partial \Psi_{AGP}}{\partial \eta_{\nu \alpha}} \right|_{\eta =0}
           = N/2 \Psi_{\nu}^+ \prod^{N/2-1} \Psi_{\alpha}^+ | vac \rangle .
\eeq

In order to see if the necessary and sufficient conditions of the exact wave function is reproduced we will first examine the effect of $\Psi_{\nu}^{+} \Psi_{\mu}^{-}$ on the AGP
\beqa
\Psi_{\nu}^{+} \Psi_{\mu}^{-} 
            \prod^{N/2} \Psi_{\alpha}^{+} | vac \rangle &=&
\Psi_{\nu}^{+} (\delta_{\mu \alpha} + R_{\mu \alpha} + \Psi_{\alpha}^{+} \Psi_{\mu}^{-}) \prod^{N/2 - 1} \Psi_{\alpha}^{+} | vac \rangle \nonumber \\
&=& \delta_{\mu \alpha} \Psi_{\nu}^{+} \prod^{N/2 - 1} \Psi_{\alpha}^{+} | vac \rangle 
+ \Psi_{\nu}^{+} (\Psi_{\alpha}^{+} R_{\mu \alpha} + \Psi_{\mu \alpha \alpha}^{+}) \prod^{N/2 - 2} \Psi_{\alpha}^{+} | vac \rangle \nonumber \\
& &+ \Psi_{\nu}^{+} \Psi_{\alpha}^+ (\Psi_{\alpha}^{+} \Psi_{\mu}^{-} + \delta_{\mu \alpha} + R_{\mu \alpha} ) \prod^{N/2 - 2} \Psi_{\alpha}^{+} | vac \rangle \nonumber \\
&=& \delta_{\mu \alpha} \frac{N}{2} \Psi_{\nu}^{+} \prod^{N/2 - 1} \Psi_{\alpha}^{+} | vac \rangle 
\label{psipsionagp}
+ \frac{1}{8}(N^2 -2N ) \Psi_{\nu}^{+} \Psi_{\mu \alpha \alpha}^{+} \prod^{N/2 - 2} \Psi_{\alpha}^{+} | vac \rangle.
\eeqa
In Eq. \ref{psipsionagp} two terms appears. The first term is a simple replacement of $\Psi_{\alpha}^{+}$ with $\Psi_{\nu}^{+}$ which also comes from the rotation of the geminals in Eq.\ref{1dev}. In the second term two geminals are replaced meaning that the variation in the form of $\Psi_{\nu}^{+} \Psi_{\mu}^{-}$ on the AGP cannot be considered a simple rotation of the geminals as in Eq.\ref{1dev}.  
We notice that the second term will disappear for $N=2$.

From Eq. \ref{1dev} we see that the energy of the AGP wave function can be minimized by varying the geminal $\Psi_{\alpha}^{+}$ but from Eq. \ref{psipsionagp} it follows that the application of $\Psi_{\nu}^{+} \Psi_{\mu}^{-}$ on the AGP wave function the minimization does not reproduce the stricter conditions for the exact wave function in Eq. \ref{strict_gem}. The AGP wave function is therefore not invariant to variations in the form of $\Psi_{\nu}^{+} \Psi_{\mu}^{-}$ since these geminal variations rotate the AGP wave function to a non-AGP wave function. 

While the AGP does not fulfill the necessary and sufficient conditions for the exact wave function it is still a relatively inexpensive correlation method which describe basic correlation.

\subsection{The antisymmetrized product of geminals (APG)}

The APG is a tensor product of $N/2$ different geminals and has been used extensively in both quantum chemistry and nuclear physics\cite{Dyson1956,BELIAEV1962582,JANSSEN1971145,10.1143/PTP.57.1554,PhysRevC.52.1394,DOBACZEWSKI1981213,RevModPhys.63.375,RevModPhys.84.711} though often in connection with the strong orthogonality. The antisymmetrized product of Strongly Orthogonal geminals (APSG)\cite{apsg1,apsg2}, using Arai's theorem\cite{arai_theorem,arai_theorem2}, 
gives a compact wave function with a reasonable accuracy.\cite{doi:10.1021/jp502127v}

The APG wave function is built from the tensor product of $N/2$ different geminals\cite{silver1,silver2,silvern(n-1)/2,nicely1}
\beq
\label{apg}
\Psi_{APG} = \prod_{\alpha}^{N/2} \Psi_{\alpha}^+ | vac \rangle.
\eeq

\subsubsection{Variations of the APG}
\label{varapg}

Varying the geminal in the APG
\beqa
\Psi_{APG} &=& \prod^{N/2}_{\alpha} \tilde \Psi_{\alpha}^+ | vac \rangle \nonumber \\
\label{APG_ansatz}
           &=& \prod^{N/2}_{\alpha} ( \Psi_{\alpha}^+ + \sum_{\nu\neq \alpha} \eta_{\nu \alpha} \Psi_{\nu}^+) | vac \rangle
\eeqa
gives the APG wave function in the transformed geminal basis $\tilde \Psi_{\alpha}^+$. Taking the first derivative of the transformed wave function with respect to the variational coefficients $\eta_{\nu \alpha}$ should reproduce the stationary conditions
\beq
\label{1dev_apg}
\left. \frac{\partial \Psi_{APG}}{\partial \eta_{\nu \alpha}} \right|_{\eta =0}
           = \Psi_{\nu}^+ \prod^{N/2-1}_{\beta \neq \alpha} \Psi_{\beta}^+ | vac \rangle.
\eeq

In order to see if the necessary and sufficient conditions of the exact wave function is reproduced we will first examine the effect of $\Psi_{\nu}^{+} \Psi_{\mu}^{-}$ on the APG
\beqa
\Psi_{\nu}^{+} \Psi_{\mu}^{-} \prod^{N/2}_{\alpha} \Psi_{\alpha}^+ | vac \rangle &=&
\Psi_{\nu}^{+} (\delta_{\mu \alpha} + R_{\mu \alpha} +\Psi_{\alpha}^{+} \Psi_{\mu}^{-}) \prod^{N/2 - 1}_{\beta \neq \alpha} \Psi_{\beta}^{+} | vac \rangle \nonumber \\
&=& \delta_{\mu \alpha} \Psi_{\nu}^{+} \prod^{N/2 - 1}_{\beta \neq \alpha} \Psi_{\beta}^{+} | vac \rangle 
+ \Psi_{\nu}^{+} (\Psi_{\beta}^{+} R_{\mu \alpha} + \Psi_{\mu \alpha \beta}^{+}) \prod^{N/2 - 2}_{\gamma \neq \alpha,\beta} \Psi_{\gamma}^{+} | vac \rangle \nonumber \\
& &+ \Psi_{\nu}^{+} \Psi_{\alpha}^+ (\Psi_{\beta}^{+} \Psi_{\mu}^{-} + \delta_{\mu \beta} + R_{\mu \beta} ) \prod^{N/2 - 2}_{\gamma \neq \alpha,\beta} \Psi_{\gamma}^{+} | vac \rangle \nonumber \\
&=& \Psi_{\nu}^{+} \sum_{\alpha} \delta_{\mu \alpha} \prod^{N/2 - 1}_{\beta \neq \alpha} \Psi_{\beta}^{+} | vac \rangle 
\label{psipsionapg}
+ \Psi_{\nu}^{+} \sum_{\alpha < \beta } \Psi_{\mu \alpha \beta}^{+} \prod^{N/2 - 2}_{\gamma \neq \alpha,\beta} \Psi_{\gamma}^{+} | vac \rangle
\eeqa
where we have assumed a canonical ordering of the geminals. Again for the APG the first term is a simple replacement of any geminal with index $\mu$ with $\Psi_{\nu}^{+}$ corresponding to a rotation of the geminal. In the second term two geminals are replaced when applying $\Psi_{\nu}^{+} \Psi_{\mu}^{-}$ on the APG.

Just like the AGP can the energy of the APG wave function can be minimized by varying the occupied geminals but from Eq. \ref{psipsionapg} it follows that the application of $\Psi_{\nu}^{+} \Psi_{\mu}^{-}$ on the APG wave function the minimization does not appear to reproduce the stricter conditions for the exact wave function in Eq. \ref{strict_gem} unless the linear combination in the second term is zero or can be added to the first term as part of a rotation. The APG wave function therefore does not appear to be invariant to geminal variations from $\Psi_{\nu}^{+} \Psi_{\mu}^{-}$ since these geminal variations appear to rotate the APG wave function to a non-APG wave function.

\subsection{The full geminal product (FGP)}

The FGP ansatz is simply
\beq
\label{fgp}
\Psi_{FGP} = \sum_{\nu_1 \leq \nu_2 \leq \ldots \nu_{N/2}} C_{\nu_1  \nu_2 \ldots \nu_{N/2}} \Psi_{\nu_1}^{+} \Psi_{\nu_2}^{+} \ldots \Psi_{\nu_{N/2}}^{+} | vac \rangle = \sum_p C_p \Psi_p
\eeq
the sum of all possible tensor products of the geminals with the expansion coefficient $C$. Here $\Psi_p$ is the $p$ tensor product with the corresponding expansion coefficient $C_p$. Since the geminals commute, as seen in Eq. \ref{commutation_op}, the tensor product of geminals in Eq. \ref{fgp} is written in canonical ordering in order to avoid duplicate products.

\subsubsection{Variations of the FGP}
\label{varfgp}

For the FGP we will follow a proof for the FCI where the configurations are swapped with tensor products of geminals.\cite{nak1}
From the stationary conditions in Eq. \ref{variation} we find
\beq
\langle \Psi_{FGP} | (\hat H - E) | \Psi_p \rangle = 0
\eeq
from the derivative of Eq. \ref{fgp}. Since $\Psi_{FGP}$ is spanned by all possible geminal tensor products $\Psi_{FGP}$ is complete and any function $\Psi_{\nu}^{+} \Psi_{\mu}^{-} \Psi_p$ can therefore be written as $\Psi_k$ which is a linear combination of the functions spanned by $\Psi_{FGP}$. Because $\Psi_k$ can be expanded in the set of $\Psi_p$ we can therefore write
\beq
\langle \Psi_{FGP} | (\hat H - E)  \Psi_{\nu}^{+} \Psi_{\mu}^{-} |\Psi_k \rangle = 0
\eeq
from which follows that the FGP is exact 
\beq
\label{fgp_exact}
\langle \Psi_{FGP} | (\hat H - E)  \Psi_{\nu}^{+} \Psi_{\mu}^{-} |\Psi_{FGP} \rangle = 0.
\eeq

That a complete product of geminals is exact is not surprising since R{\o}eggen have shown that the extended geminal model is exact.\cite{inge1,inge5,inge6,inge_book}

The FGP not only fulfill the necessary and sufficient conditions for the exact wave function in Eq. \ref{double_g}, as expressed in Eq. \ref{fgp_exact}, but also higher order geminal products
\beqa
\label{fgp_exact_high}
\langle \Psi_{FGP} | (\hat H - E)  \Psi_{\nu}^{+} \Psi_{\xi}^{+} \Psi_{\omicron}^{-} \Psi_{\mu}^{-} |\Psi_{FGP} \rangle &=& 0 \nonumber \\
\langle \Psi_{FGP} | (\hat H - E)  \Psi_{\nu}^{+} \Psi_{\xi}^{+} \Psi_{\pi}^{+} \Psi_{\rho}^{-} \Psi_{\omicron}^{-} \Psi_{\mu}^{-} |\Psi_{FGP} \rangle &=& 0 \nonumber \\
\ldots
\eeqa
since these higher products also can be expanded in the set of $\Psi_p$. That the FGP also fulfill higher order products is simply due to having a complete set of products, exactly as seen for the FCI with a complete set of configurations.\cite{nak1}

The FGP does not appear to have any numerical advantage over the FCI when all terms in the expansion are included. A truncated version of the FGP could be significantly more compact than CI since the first term in the FGP is of APG quality.

\section{Summary and prospects}

We have here shown the transformation from a one-particle basis to a geminal basis by defining a unit geminal basis. In this way both the wave function and Hamiltonian are written in the geminal basis. It is shown that the necessary and sufficient conditions of the exact wave function\cite{nak1,nak2} can be written as a Brillouin theorem of geminals which is consistent with the generalized Brillouin theorem derived previously.\cite{Lasse_Brillouin}

A significant amount of space have been dedicated to the Lie algebra of the geminals in order to compare the geminal rotations with the effect of, what in a one-particle basis would be called, the replacement operator $\Psi_{\nu}^{+} \Psi_{\mu}^{-}$. We here show that $\Psi_{\nu}^{+} \Psi_{\mu}^{-}$ introduce primary and secondary rotations. The primary rotations reproduce the regular rotations of the geminals but the secondary rotations rotate two geminals in the reference at the same time where one of the geminals is a linear combination of multiple geminals. The origin of the secondary rotations are the composite boson behaviour of the geminals where the commutation between the creator and annihilator geminals produce an additional term $R$, which does not commute with any other operators. 

We have gone through the simplest ans{\"a}tze for a wave function in a geminal basis and compared the variation of the geminals to the exact stationary conditions. For the antisymmetrized geminal power (AGP) we show that the secondary rotations from $\Psi_{\nu}^{+} \Psi_{\mu}^{-}$ rotate the AGP away from a pure AGP wave function. The AGP is therefore not invariant to variations from $\Psi_{\nu}^{+} \Psi_{\mu}^{-}$. The antisymmetrized product of geminals (APG) appears to be rotated away from a pure APG though this is less obvious than for the AGP.

The full geminal product (FGP), which is a tensor product of all possible geminals, is shown to be exact. Using all terms in the FGP does not give any advantage over the full configuration interaction (FCI), however, a truncated version of the FGP could give a significantly more compact representation of the wave function since the lowest order of the FGP can be chosen to be the AGP.

The clear advantage of the these general geminal wave function ans{\"a}tze is the ability of including all determinants in the FCI in the wave function at the same time without the exponential scaling. The coefficients in front of all determinants in the geminal wave functions of course cannot vary freely, like in the FCI, but are constrained by the wave function ansatz. Despite these constraints this should make geminal wave functions more agnostic with respect to static and dynamic correlation and simply include the most important correlation, a feat that has proven very difficult for wave functions expanded in a one-particle basis set.

\begin{appendix}


 \section{Additional geminal and orthogonality relations summarized}
\label{gemalg}

The relations summarized here are considered 
obvious though still useful.
The relations show the effect of different operators
working on the vacuum state
\beqa
R_{\mu \nu} | vac \rangle &=& 0 \\
Q_{\mu \nu} | vac \rangle &=& \delta_{\mu \nu} | vac \rangle \\
\Psi_{\mu}^- | vac \rangle &=& 0 \\
(\Psi_{\mu}^+ \Psi_{\nu}^+) | vac \rangle &\neq& 0 \\
(\Psi_{\mu}^+ \Psi_{\mu}^+) | vac \rangle &\neq& 0 \\
(\Psi_{\mu}^- \Psi_{\nu}^+) | vac \rangle &=& \delta_{\mu \nu} | vac \rangle
\eeqa
and simple commutation relations with an orbital
\beq
[\Psi_{\mu}^+,a_{p}^{\dagger}] = [\Psi_{\mu}^- , a_{p}] = 0.
\eeq

\section{The necessary and sufficient conditions for the exact wave function}
\label{sec:necessary}

We here start from the time-independent many-body electronic wave function projected onto some one-particle basis. In the language of second quantization this projection gives the familiar electronic Hamiltonian $\hat H$ in second quantization
\beq
\label{hamiltonian}
\hat H = \hat f + \hat g
       = \sum_{pq} f_{pq} a_p^{\dagger} a_q
       + \sum_{pqrs} g_{pqrs} a_p^{\dagger} a_r^{\dagger} a_s a_q ,
\eeq
where $\hat f$ and $\hat g$ describe the one- and two-electron interaction, respectively.
The eigenfunctions and eigenvalues for the Hamiltonian in Eq. \ref{hamiltonian} 
\beq
\hat H \Psi = E \Psi
\eeq
are then the solutions to the projected time-independent many-body electronic equation. 

By rewriting the one-electron operator and adding this to the two-electron operator
\beq
\label{allH}
\hat H = \sum_{pqrs} (\frac{f_{pq} \delta_{rs}}{N-1} + g_{pqrs} ) a_p^{\dagger} a_r^{\dagger} a_s a_q = \sum_{pqrs} \omega_{pqrs} a_p^{\dagger} a_r^{\dagger} a_s a_q
\eeq
it is straight forward to derive the necessary and sufficient conditions for the exact wave function\cite{nak1,nak2,Lasse_Brillouin}
\beq
\label{double}
0 = \langle \Psi | (\hat H - E) \omega_{pqrs} a_p^{\dagger} a_r^{\dagger} a_s a_q |  \Psi \rangle .
\eeq
In this way the necessary and sufficient conditions or stationary conditions for the exact wave function are written as a generalized Brillouin theorem.\cite{Lasse_Brillouin}

\section{Rotations, Hamiltonian and parameters in a geminal basis}
\label{appendix_ham}

We will here define the simplest possible geminal basis and show the effect of orbital and geminal rotations for this simple geminal basis and other bases. From the geminal rotations we will write down the Hamiltonian in an arbitrary geminal basis. Finally the transformation to a natural geminal is discussed. 

\subsection{A simple geminal basis}
\label{ssec:simple}

Starting from a one-particle basis set we will introduce the simplest possible geminal basis, for which will use the labels i,j,$\ldots$, 
\beq
\label{simple_geminal}
\Psi_{i}^+ = \frac{1}{2} ( C_{pq} a^{\dagger}_p a^{\dagger}_q + C_{qp} a^{\dagger}_q a^{\dagger}_p) \qquad
 C_{pq} = \left\{
\begin{array}{rl}
1  &\text{if } p > q, \\
-1 &\text{if } p < q, \\
0  &\text{if } p = q, 
\end{array}
\right.
\eeq
where we join the two spin-orbital indices to a single label in the geminals basis. We here define an annihilation geminal from the Hermitian adjoint of Eq. \ref{simple_geminal}
\beq
\label{simple_geminala}
 \Psi_{i}^- = \frac{1}{2} ( C_{pq} a_q a_p +  C_{qp} a_p a_q) \qquad
 C_{pq} = \left\{
\begin{array}{rl}
1  &\text{if } p > q, \\
-1 &\text{if } p < q, \\
0  &\text{if } p = q.
\end{array}
\right.
\eeq
With this definition a complete orthonormal geminal basis is defined, according to the Frobenius inner product, where the number of geminals in the basis will be $m(m-1)/2$ for $m$ spin-orbitals since we enforce skew-symmetry of the coefficient matrix. This simple basis can be considered a unit basis for geminals.

Any geminal from the basis defined in Sec. \ref{sec:defgem} can now be expanded in this simple basis
\beq
\label{geminal_simple}
\Psi_{\mu}^{+} = \sum_{i} c_{\mu i}  \Psi_{i}^+ = \sum_{pq} C_{pq}^{\mu} a^{\dagger}_p a^{\dagger}_q
\eeq
and likewise for the annihilation geminal
\beq
\label{geminala_simple}
\Psi_{\mu}^{-} = \sum_{i} c_{\mu i}  \Psi_{i}^- =  \sum_{pq} C_{pq}^{\mu} a_q a_p.
\eeq

\subsection{Orbital and geminal rotations}

The simple geminal basis as shown in Eq. \ref{geminal_simple} will of course change with the rotation of one-electron functions. We will here therefore relate two simple geminal bases where the one-particle function have been rotated. In the rotated basis, denoted with a bar, the simple geminal can be written as
\beq
\label{simple_geminal_rot}
\bar \Psi_{i}^+ = \frac{1}{2} ( C_{pq} \bar a^{\dagger}_p \bar a^{\dagger}_q +  C_{qp} \bar a^{\dagger}_q \bar a^{\dagger}_p).
\eeq
Since the orbitals are related by some unitary transformation the pair of orbitals are related as
\beq
\bar a^{\dagger}_p \bar a^{\dagger}_q = \sum_{rs} U_{pr} a^{\dagger}_r U_{qs} a^{\dagger}_s
\eeq
and therefore the also simple geminals
\beq
\bar \Psi_{i}^+ = \sum_j U_{ij} \Psi_{j}^+
\eeq
where every column of $U_{ij}$ is a vectorized coefficient matrix consistent with the definition of the Frobenius inner product.

Since the Frobenius inner product is used on the coefficient matrix to create an orthonormal geminal basis the expansion coefficients in Eq. \ref{geminal_simple} will also be a simple unitary transformation to a new basis
\beq
\Psi_{\mu}^{+} = \sum_{i} U_{\mu i}  \Psi_{i}^+ = \sum_{i} \bar U_{\mu i}  \bar \Psi_{i}^+ = \sum_{\nu} \bar U_{\mu \nu} \bar \Psi_{\nu}^{+}
\eeq
irrespective of the choice of one-particle basis.

\subsection{Hamiltonian in a geminal basis}

Knowing the transformation from the one-particle basis to the simple geminal basis and geminals, as shown in Sec. \ref{sec:defgem}, writing the Hamiltonian from Eq. \ref{allH} in any geminal basis is straight forward
\beq
\label{H_in_gem_basis}
\sum_{pqrs} \omega_{pqrs} a^{\dagger}_p a^{\dagger}_q a_r a_s = \sum_{ij} \omega_{ij} \Psi_{i}^+ \Psi_{j}^- = \sum_{ij} \omega_{ij} \sum_{\mu} U_{i \mu} \Psi_{\mu}^+ \sum_{\nu} U_{i \nu} \Psi_{\nu}^- = \sum_{\mu \nu} \omega_{\mu \nu} \Psi_{\mu}^+ \Psi_{\nu}^-.
\eeq
In Eq. \ref{H_in_gem_basis_a} every pair index have also been joined for $\omega$ where the Hamiltonian is in a geminal basis.

\subsection{Natural geminal}
\label{ssec:natural}

Geminals are more often used in the natural form\cite{silver1}
\beq
\label{geminald}
\Psi_{\mu}^{+} =  \frac{1}{2} \sum_{p}^m C_{p} a^{\dagger}_{p \alpha} a^{\dagger}_{p \beta}
\eeq
where $C_{p}$ still is an element in a skew-symmetric coefficient matrix and the sum now only is the orbital pairs in a spin-restricted basis. The general and natural geminal can be related through an orbital transformation though this orbital transformation will in general destroy both the spatial and spin pairing of the orbitals so the nice pairing seen in Eq. \ref{geminald} cannot be expected from such a transformation.\cite{Shull56,ortiz81} Eq. \ref{geminal} can only be transformed into the natural geminal as written in Eq. \ref{geminald} for a two-electron problem with low spin. 

The exact orbital transformation to the natural geminal geminal cannot be known \maps but can only be constructed \mapos or iteratively. In a geminal basis the orbital rotation to the natural form of a geminal will in general be different for each geminal in the geminal basis. Since a maximum of $m$ geminals of the $m(m-1)/2$ geminals in a geminal basis with $m$ one-particle functions can be natural at the same time there is no advantage in trying make some geminals in a basis natural when the entire geminal basis is used.


 
\end{appendix}

\bibliographystyle{unsrt}
\bibliography{main}

\end{document}